\documentstyle[aps]{revtex}
\tighten

\def\be{\begin{equation}}
\def\ee{\end{equation}}
\def\beq{\begin{eqnarray}}
\def\eeq{\end{eqnarray}}
\def\n{\nonumber}

\begin{document}


\title{Relaxational effects in radiating stellar collapse}

\author{
Megan Govender$^{1}$, Roy Maartens$^{2}$ and
Sunil D Maharaj$^{3}$
}

\address{~}

\address{$^1$Department of Physics, Technikon Natal,
PO Box~953, Durban~4000, South Africa}

\address{$^2$School of Computer Science and Mathematics,
Portsmouth University, 
Portsmouth~PO1~2EG, England}

\address{$^3$Department of Mathematics, University of Natal,
Durban~4041, South Africa}

\address{~}

\date{30 September 1998}

\maketitle

\begin{abstract}

Relaxational effects in stellar heat transport
can in many cases be significant. Relativistic Fourier-Eckart
theory is inherently quasi-stationary, and cannot incorporate these
effects. The effects are naturally accounted for in
causal relativistic thermodynamics, which provides an improved
approximation to kinetic theory. 
Recent results,
based on perturbations of a static star,
 show that relaxation effects can produce a
significant increase in the central temperature and
temperature gradient for a given luminosity.
We use a simple stellar model
that allows for non-perturbative deviations from staticity,
and confirms qualitatively the predictions of the perturbative models.

\end{abstract}

~\\



\section*{1 Introduction}

The formation of compact astrophysical objects such as
neutron stars and
black holes is usually preceded by a period of radiative 
gravitational
collapse in which neutrinos and photons carry energy to the exterior
spacetime. The problem of radiative collapse was originally
addressed by Oppenheimer and Snyder (1939), who investigated a
collapsing dust sphere.
Their idealised treatment
has subsequently been improved via
more physically realistic thermodynamical models.

Most work has been based on the standard Fourier-Eckart
theory of heat transport (see, e.g., Grammenos, 1994). However, this
theory is non-hyperbolic, and thus
suffers from the pathologies of infinite
propagation speed of thermal signals, and instability of all
equilibrium states (Israel and Stewart, 1979, Hiscock and Lindblom,
1983). The causal (hyperbolic) generalization of standard
thermodynamics, due to Israel and Stewart (1979),
is based on an improved
kinetic theory approximation, and avoids these pathologies.
The causal behaviour results essentially from treating the
heat flux as an independent dynamical quantity,
on the same footing as
the equilibrium quantities. As a consequence, non-quasi-stationary
and transient relaxational effects are incorporated,
expressed via the appearance of
a time derivative of heat flux in the
transport equation.
By contrast, the standard quasi-stationary transport equation
contains no time derivative; it is an algebraic relation for
heat flux, incapable of accounting for relaxational effects.

In relativistic stellar
collapse, strong deviations from quasi-stationary heat transport
can occur, so that
relaxation effects can be significant, as
argued by Di Prisco et al. (1996) and Herrera and Santos (1997).
Recently, a number of radiating collapse models with causal heat
transport has been developed (Martinez, 1996,
Di Prisco et al., 1996, Herrera and Santos, 1997, Herrera et al.,
1997,
Govender et al., 1998,
Herrera and Martinez, 1998).\footnote{
See also
Triginer and Pavon (1996) and Maartens et al. (1998) for
cosmological models with causal heat flux.
}

These models show that the causal theory can predict
significantly different results
when the relaxation time-scale is sufficiently large
relative to other time-scales. Typically, for a given luminosity,
causal corrections lead 
to a greater temperature gradient, and thus
greater central temperature. In other words, non-quasi-stationary
heat transport implies that a collapsing star with a given
observed luminosity is hotter, and with greater temperature
gradient, than predicted by quasi-stationary heat transport.

The role played by the relaxation time during radiative
gravitational collapse was investigated by Herrera and
Santos (1997), assuming that the spacetime metric
is close to a static (and non-radiating) initial metric.
Perturbation of the initial configuration is assumed to
produce collapse without shearing.
In this paper, we use a very simple
shear-free collapse model with the
same static limit as in Herrera and Santos (1997), but which
allows for large deviations from the initial static metric.
Although these deviations are very simple in form, they do
allow us to provide a
non-perturbative
complementary approach to the perturbative model
of Herrera and Santos. We are able to find explicit analytic
forms for the temperature for a simple relaxation time, using
the methods of our previous work (Govender et al., 1998).
We also consider the
case of small relaxational effects, to find a perturbative
solution for the temperature (without assuming that the metric
is close to static). Our results confirm the increase of
temperature and temperature gradient for a given luminosity.

\section*{2 Simple model}

A shear-free isotropic stellar fluid has metric
\[
ds^2=-A^2(t,r)dt^2+B^2(t,r)\left[dr^2+r^2d\Omega^2\right]\,,
\]
in comoving coordinates, where $d\Omega^2$ is the metric of the
unit 2-sphere. The fluid 4-velocity is $u^a=A^{-1}\delta^a{}_0$,
and its volume rate of expansion $\Theta=\nabla_au^a$ is
\be
\Theta={3\over A}{\dot{B}\over B}\,,
\label{exp}\ee
where an overdot denotes $\partial/\partial t$. Since the shear
rate $\sigma_{ab}$ vanishes, the expansion rates in the radial
and tangential directions are the same, and equal to $\dot{B}/(BA)$.
If the star is collapsing, then of course $\dot{B}<0$.

Modelling the star as a perfect fluid with heat conduction (where
the heat flux is provided principally by neutrinos), the 
energy-momentum tensor is
\be \label{aaa1}
T_{ab} = (\rho + p)u_au_b + pg_{ab} + q_au_b + q_bu_a\,,
\ee
where $\rho$ is the energy density, $p$ is the pressure, and 
$q_a=qn_a$ is the heat flux, with $n^a$ a unit radial vector.

Herrera and Santos (1997) assumed a shear-free perturbed metric
of the form
\be \label{x1}
ds^2 = - \left[{A_0}(r) + \epsilon a(r)Y(t)\right]^2dt^2 +
\left[{B_0}(r) + \epsilon b(r)Y(t)\right]^2\left[dr^2 +
r^2 d\Omega^2\right]\,, \label{a14b}
\ee
where $A_0$ and $B_0$ describe the initial static star,
and $\epsilon\ll1$ is the perturbation parameter.
It follows that (Bonnor et al., 1989)
\begin{eqnarray*}
{\rho}(t,r) &=& {\rho_0}(r) + \epsilon E(t,r) \,,\\ 
 p(t,r) &=& p_0(r) + \epsilon P(t,r)\,,\\
 q(t,r) &=& \epsilon Q(t,r)\,,
\end{eqnarray*}
where subscript zero denotes equilibrium quantities.
The dynamical stability of this particular model has been studied
extensively in the past by
Chan et al. (1989) and Herrera et al. (1989), but a systematic
thermodyanamical treatment of the heat flux and temperature gradient
was not carried out.

Here we consider an alternative approach to
that of Herrera and Santos (1997), where we use a simple
metric that accomodates large (though restricted) deviations
from staticity, and that leads to simple analytic expressions
for the important quantities.
We use the model first proposed by de Oliveira et al.
(1985, 1986): 
\be
ds^2 = -  {A_0}^2(r)dt^2 + {B_0}^2(r)f^2(t)\left[dr^2 + 
r^2d\Omega^2 \right]\,. \label{a1a}
\ee
It follows from equations (\ref{exp}) and (\ref{a1a})
that the expansion rate is
\be
\Theta={3\over A_0}{\dot{f}\over f}\,.
\label{exp'}\ee
Einstein's field equations for the metric (\ref{a1a})
and the energy-momentum tensor (\ref{aaa1}) give
(Bonnor et al., 1989)
\beq     \label{a4}
{\rho} &=& \displaystyle\frac{1}{f^2} \left( \rho_0 +
\displaystyle\frac{3}{{A_0}^2}{\dot{f}}^2 \right)\,, \label{a4a} \\ 
p &=&  \displaystyle\frac{1}{f^2} \left[ p_0 -
\displaystyle\frac{1}{{A_0}^2}(2f{\ddot{f}} + {\dot{f}}^2) \right]\,,
\label{a4b} \\ 
q &=&
-\displaystyle\frac{2{A_0}^{\prime}}
{{A_0}^2{B_0}}\displaystyle\frac{\dot{f}}{f^2}\,,
\label{a4c}
\eeq
where a prime denotes $\partial/\partial r$, and the pressure isotropy
condition
\be \label{a2a}
\left(\displaystyle\frac{{A_0}^{\prime}}{A_0} +
\displaystyle\frac{{B_0}^{\prime}}{B_0}\right)^{\prime} -
\left(\displaystyle\frac{{A_0}^{\prime}}{A_0} +
\displaystyle\frac{{B_0}^{\prime}}{B_0}\right)^2 -
\displaystyle\frac{1}{r}\left(\displaystyle\frac{{A_0}^{\prime}}{A_0} 
+
\displaystyle\frac{{B_0}^{\prime}}{B_0}\right) +
2\left(\displaystyle\frac{{A_0}^{\prime}}{A_0}\right)^2 = 0\,.
\ee
In equations (\ref{a4a}) and (\ref{a4b}) we have introduced
\beq \label{aa2}
{\rho}_0 &=& -\displaystyle\frac{1}{{B_0}^2}\left[
2\left(\displaystyle\frac{{B_0}^{\prime}}{B_0} \right)^{\prime} +
\left(\displaystyle\frac{{B_0}^{\prime}}{B_0} \right)^2
+ \displaystyle\frac{4}{r}\displaystyle\frac{{B_0}^{\prime}}{B_0}
\right] \,,\label{aa2a}  \\ 
p_0 &=&
\displaystyle\frac{1}{{B_0}^2}
\left[\left(\displaystyle\frac{{B_0}^{\prime}}{B_0
} \right)^2
+ 2 \displaystyle\frac{{A_0}^{\prime}}{A_0}
\displaystyle\frac{{B_0}^{\prime}}{B_0}+ 
\displaystyle\frac{2}{r}\left(\displaystyle\frac{{A_0}^{\prime}}{A_0} 
+ \displaystyle\frac{{B_0}^{\prime}}{B_0}
\right)\right]\,,  \label{aa2b}
\eeq
which define the energy density and pressure of the initial static
configuration before the onset of collapse. 
The functions $A_{0}(r)$ and $B_{0}(r)$ must satisfy the isotropy
condition (\ref{a2a})
in order to obtain a solution to the Einstein field
equations with isotropic pressure.
Failure to check this condition can invalidate a
solution (see e.g. Govinder et al., 1998).
 Equations (\ref{a4a})--(\ref{a4c}) together with
(\ref{aa2a}) and (\ref{aa2b}) describe a radiating 
shear-free star
with an initial static configuration.

The interior metric must be matched to the outgoing Vaidya metric
across a time-like hypersurface $\Sigma$, described by
$r = r_{_\Sigma} =$ constant in comoving coordinates.
The Vaidya (1953)
metric is
\[
ds^2 = -\left[1 - \frac{2m(v)}{R}\right]dv^2 - 2dvdR +
R^2d\Omega^2\,,
\]
where $m(v)$ is the mass function. 
The static solution, before the onset of collapse at $t = -\infty$,
matches the exterior Schwarzschild spacetime across 
$\Sigma$.
In the static limit, the
pressure $p_0$ vanishes at the boundary:
$(p_0)_{_\Sigma} = 0$.
At late times the model becomes nonstatic and the 
pressure $(p)_{_\Sigma}$ is 
nonzero in general
because of the presence of heat flow (Santos, 1985).
The junction conditions resulting from the matching of (\ref{a1a}) to
the outgoing Vaidya metric yield
\beq  \label{a5}
(R)_{_\Sigma} &=&(rB_{0}f)_{_\Sigma} \,, \label{a5a}  \\  
(p)_{_\Sigma} &=& (q)_{_\Sigma}\,,  \label{a5b}  \\ 
\left[ r(rB_{0}f)^\prime \right]_{_\Sigma}
&=&
\left[\dot{v}(R - 2m) + \dot{{R}}{R}
\right]_{_\Sigma} \,, \label{a5c} \\  
m(v) &=& \left( \displaystyle\frac{r^3{B_0}^3f{\dot{f}}^2}{2{A_0}^2}
- r^2{B_0^{\prime}}f - 
\displaystyle\frac{r^3{{B_0}^{\prime}}^2f}{2B_0} \right)_{_\Sigma} \,.
\label{a5d}
\eeq
The proper radius of the star is 
\be \label{ss1}
r_{\rm p}(t) = \int_{0}^{r_{_\Sigma}}B_0(r) f(t)dr\,.
\ee
Using the junction condition (\ref{a5b}) together with (\ref{a4b}) 
and
(\ref{a4c}), and taking into account $(p_{0})_{_\Sigma} = 0$, we 
obtain the evolution equation
\be \label{a6}
2f{\ddot{f}} + {\dot{f}}^2 - 2a{\dot{f}} = 0 \,.
\ee
The constant
\be  \label{a7}
a = \left(\displaystyle\frac{{A_0}^{\prime}}{B_0} \right)_{_\Sigma}
\ee
is positive because the static solution $(A_{0}, B_{0})$
must match with the exterior Schwarzschild metric.
A first integral of (\ref{a6}) is given by
\[
\dot{f} = -2a\left(\displaystyle\frac{b}{\sqrt{f}} - 1\right)\,,
\]
where the constant of integration is $-2ab$. We choose $b = 1$ so 
that
$\dot f \rightarrow 0$ when $f \rightarrow 1$ in the static limit.
Thus
\be \label{a8}
\dot{f} = -2\left({{A_0}^\prime\over B_0}\right)_{_\Sigma}
\left(\displaystyle\frac{1}{\sqrt{f}} - 1\right)\,.
\ee
Note that $p_{_\Sigma}$ is nonnegative, so on using the result
$(p_{0})_{_\Sigma} = 0$, together with equations
(\ref{a4b}) and (\ref{a6}), we obtain
$\dot{f} \leq  0$. 
By equations (\ref{exp'}) or (\ref{ss1}),
this implies that the only possible evolution is
contraction. Also, equation 
(\ref{a8}) gives $|f|\leq1$, and without loss of generality we
assume $f$ is non-negative, so that
\be \label{www}
0 \leq f(t) \leq 1 \,.
\ee
Integrating equation (\ref{a8}) we obtain
\be \label{a9}
t = \displaystyle\frac{1}{a}
\left[\displaystyle\frac{1}{2}f + \sqrt{f} + \ln{\left(1 -
\sqrt{f} \right)} \right] \,,
\ee
where the constant of integration has been absorbed by rescaling 
$t \rightarrow t + {\mbox{constant}}$.
This shows that $f$ decreases
monotonically from  $1$ at $ t = -\infty$ to zero at $ t = 0$,
where a physical singularity is encountered, 
since $\rho\rightarrow\infty$ by equation (\ref{a4a}).
From equation (\ref{a5d}) we can obtain the simpler
expression
\be \label{mee}
m(v) = \left[\frac{2a^2r^3{B_0}^3}{{A_0}^2}\left(1 - \sqrt{f}
\right)^2 + m_0
f\right]_{_\Sigma}\,,
\ee
where the initial mass $m_0$ of the star is given by
\be \label{a10}
m_0 = -\left(r^2{B_0}^{\prime} +
r^3\displaystyle\frac{{{B_0}^{\prime}}^2}{2B_0}\right)_{_\Sigma}\,,
\ee
since at $t = -\infty$ the model is static.
In the infinite past we must match the static perfect fluid solution 
to the exterior Schwarzschild
solution in isotropic coordinates and we obtain
\beq
\left({A_0}^{\prime}\right)_{_\Sigma} &=&
\displaystyle\frac{m_0}{{{{{\cal R}_{0}}}}^2}\left(1 +
\displaystyle\frac{m_0}{2r_{_\Sigma}}\right)^2\,,
\label{a11a} \\
\left({B_0}\right)_{_\Sigma} &=& \left(1 + 
\displaystyle\frac{m_0}{2r_{_\Sigma}}\right)^2\,,
\label{a11b}
\eeq
where
\[
{\cal R}_0 = \left(rB_0\right)_{_\Sigma}
\]
is the initial luminosity radius.
We can rewrite equation (\ref{a7}) as
\[
a = \displaystyle\frac{m_0}{{{{{\cal R}}_{0}}}^2}\,,
\]
where $m_0$ is given by (\ref{a10}).
The surface redshift $z_{_\Sigma}$ is
\be   \label{a21}
z_{_\Sigma} = \left[\frac{r(B_0 r)^{\prime}f +
2ar^2{B_0}^2{A_0}^{-1}\sqrt{f}\left(1 - \sqrt{f}\right)}{r{B_0 f} -
2m}\right]_{_\Sigma} - 1 \,.
\ee
This becomes infinite at 
\be 
r_{\rm h} = \left(\frac{2m}{B_0f}\right)_{_\Sigma} \,,
\label{hor}\ee
which locates the
horizon. By equations (\ref{a4c}) and (\ref{a8}) we get
\be  \label{ku1}
q = \frac{4a{A_0}^{\prime}}{{A_0}^2B_0}\left(
\frac{1 - \sqrt{f}}{f^2\sqrt{f}}\right) \,.
\ee
We require $q>0$, which implies that
\[
{A_0}^{\prime} >0 \,.
\]
For a physically reasonable model the energy density and
pressure must be decreasing functions of the radial coordinate.
From equations
(\ref{a4a}) and (\ref{a4b}) we can write the spatial gradients 
of the energy density and pressure as
\beq \label {st1}
{\rho}^{\prime} &=& {{\rho}_0}^{\prime}\frac{1}{f^2} -
24a^2\frac{{A_0}^{\prime}}{{A_0}^3}\frac{(1 - \sqrt{f})^2}{f^3}\,,
\label{st1a}\\ 
p^{\prime} &=& {{p}_0}^{\prime}\frac{1}{f^2} -
8a^2\frac{{A_0}^{\prime}}{{A_0}^3}\left(
\frac{1 - \sqrt{f}}{f^2\sqrt{f}}\right)\,.
\label{st1b} 
\eeq
 From these equations it follows that in order to have 
${\rho}^{\prime} < 0$ and
$p^{\prime} < 0$, we must have  ${{\rho}_0}^{\prime} < 0$ and
${p_0}^{\prime} < 0$ for the initial static configuration. It has 
been shown by Bonnor et al. (1989) that the strong energy condition
is satisfied throughout the collapse if the initial static 
configuration
satisfies the condition
\be \label{st2}
{\rho}_0 - 3p_0 \geq \frac{3a^2}{{A_0}^2}
\ee
at $r = 0$.

\section*{3 Causal heat transport}

We now investigate the evolution of the temperature profile using the 
causal
thermodynamics developed by Israel and Stewart (1979). The
causal relativistic transport equation is
\be \label{a12}
\tau h_a{}^bu^c\nabla_c
{q}_b + q_a = -\kappa \left(
h_a{}^b\nabla_b T+Tu^b\nabla_b{u}_a\right)\,,
\ee
where $h_{a}{}^b=\delta_{a}{}^b+u_a u^b$
is the projection tensor into the comoving rest space, 
$T$ is the local equilibrium
temperature, $\kappa$ is the thermal conductivity and
$\tau$ is the relaxational time-scale. Setting $\tau = 0$ in
(\ref{a12}) we obtain the non-causal Fourier-Eckart
law for quasi-stationary heat transport, which leads to
pathological behaviour in the propagation velocity of thermal
signals.

For the metrics (\ref{x1}) and (\ref{a1a}), equation (\ref{a12}) 
reduces to
\be \label{a14}
\tau{\dot q} + Aq = -\displaystyle\frac{\kappa}{B}{(AT)}^{\prime}\,.
\ee
Taking $\tau \approx 10^{-4}$ s, which corresponds to the
early stages of the evolution of a neutron star, Herrera and 
Santos (1997)
found that the magnitude of the temperature gradient was enhanced
by 10 per cent as compared to the case $\tau = 0$. For a temperature
of approximately $10^6$ K and a thermal propagation velocity
$v \approx 10^3$ cm/s, Di Prisco et al.
(1996) obtained $\tau \approx 10^2$ s.
Using this value as an upper bound on the relaxation time,
Herrera and Santos found that the
temperature gradient is five orders greater than the corresponding
non-causal case. In general, they concluded that for a
given luminosity, the associated temperature gradient is enhanced
by relaxational effects, and the enhancement grows with
increasing relaxation times. (This qualitative conclusion is
confirmed by Di Prisco et al., 1996, and Govender et al., 1998.)

In this section we 
perform a similar analysis but with the simple non-perturbative
metric (\ref{a1a}).
For a physically reasonable model, we assume that heat is
carried to the exterior spacetime by thermally generated neutrinos
having long mean free paths through the stellar core.
The thermal conductivity is then given by the radiative transfer
form (Martinez, 1996)
\be
\kappa =\gamma T^3{\tau}_{\rm c} \,,  \label{a15}
\ee
where $\gamma$ ($\geq0$) is a constant and ${\tau}_{\rm c}$ is 
the mean collision time for neutrino-matter interactions. 
These interactions are
primarily due to electron-neutrino scattering and nucleon absorption.
Following Govender et al. (1998), 
we assume a mean collision time of form
\be
\tau_{\rm c} =\left({\alpha\over\gamma}\right) T^{-\sigma} \,,
\label{a16}\ee
where $\alpha$ ($\geq 0$) and $\sigma$ ($\geq 0$) are constants. 
For thermal neutrino transport, Martinez (1996) argues that
$\sigma\approx {3\over2}$.
The mean collision time decreases with growing temperature except for
the special case $\sigma=0$, when it is constant. The special case 
of constant $\tau_{\rm c}$ was investigated by Govender et al. (1998) 
for
acceleration-free gravitational collapse.
By adopting a similar approach,
we investigate the relaxation effects
when the acceleration is nonzero.
The relaxation time-scale in causal radiative transfer (Udey and
Israel, 1982, Schweizer, 1988) 
is taken as equal to the mean collision time.
A simple generalization
\be
\tau =\left({\beta\gamma \over \alpha}\right) \tau_{\rm c} \,,
\label{a17}\ee
where $\beta$ ($\geq 0$) is a constant, allows both
for cases where the relaxation time-scale is greater than
the collision time-scale ($\beta\gamma/\alpha>1$) (Maartens
and Triginer, 1998), and for the case of perturbative deviations
from the quasi-stationary case, i.e. $\beta\gamma/\alpha\ll1$. The
degenerate case $\beta=0$ recovers the non-causal transport law.

With these assumptions, the transport equation (\ref{a14}) 
implies
\be \label{a18}
\left(A_0T\right)^{3-\sigma}\left(A_0T\right)^\prime
-2{\beta\over\alpha}\left({\ddot{f}\over f}-2{\dot{f}^2\over f^2}
\right){A_0}{A_0}^\prime\left(A_0 T\right)^{-\sigma}=
{2\over\alpha}{\dot f\over f}
{A_0}^{2-\sigma}{A_0}^{\prime} \,,
\ee
where $f(t)$ is implicitly determined by equation (\ref{a9}).
Equation (\ref{a18}) is a first order radial equation in
$A_0T$, which is solved in principle once $A_0(r)$ is specified
and boundary conditions are given to determine the time-function
of integration. 
The effective surface temperature as determined by an external 
observer
is given in general by Misner (1969);
for the metric (\ref{a1a}) this leads to
\be \label{a19}
\left(T^4\right)_{_\Sigma} = \left[\frac{1}{\xi(rB_0
f)^2}\right]_{_\Sigma}L_{\infty}\,,
\ee
where $\xi$ is a constant, and $L_\infty$ is
the total luminosity for an observer at infinity.
This can be found using
the general form in Bonnor et al. (1989):
\be \label{a20}
L_{\infty} = 
2a^2\left(\frac{{B_0}^2r^2}{{A_0}^2}\right)_{_\Sigma}\left(\frac{1
- \sqrt{f}}{\sqrt{f}}\right)\frac{1}{(1 + z_{_\Sigma})^2}\,,
\ee
where the surface redshift $z_{_\Sigma}$ is given by equation
(\ref{a21}).

Equations (\ref{a19}) and (\ref{a20}) determine the time-function
of integration of the temperature equation (\ref{a18}).
In the non-causal case $\beta=0$, the temperature equation is
readily solved. Writing $\tilde{T}$ for the non-causal temperature,
we find
\beq
\left(A_0\tilde{T}\right)^{4-\sigma} &=&{2\over\alpha}\left(
{4-\sigma\over3-\sigma}\right){\dot{f}\over f}{A_0}^{3-\sigma}+{\cal 
F}~
\mbox{ for }~\sigma\neq3,4\,, \label{roy1}\\
A_0\tilde{T} &=& {2\over\alpha}{\dot{f}\over f}\ln A_0+{\cal 
F}~\mbox{ for }~
\sigma=3\,, \label{roy2}\\
\ln\left(A_0\tilde{T}\right) &=& -{2\over\alpha}{\dot{f}\over f}
{A_0}^{-1}+{\cal F}~\mbox{ for }~\sigma=4 \,, \label{roy3}
\eeq
where ${\cal F}(t)$ is determined from the surface conditions via
equations (\ref{a21}), (\ref{a19}) and (\ref{a20}).
For example, if $\sigma=0$, then we find
\be
{\cal F}(t) = \left\{\frac{2a^2 {A_0}^4}{\xi r^2}\frac{(1 -
\sqrt{f})}{{f}^{3}\sqrt{f}}\left[\frac{rB_0f -
2m}{A_0(B_0r)^{\prime}\sqrt{f} + 2ar{B_0}^2(1 - \sqrt{f})}\right]^2 -
\frac{8}{3\alpha}{A_0}^3 \frac{\dot
f}{f}\right\}_{_\Sigma}\,.
\label{roy4}\ee

The temperature equation (\ref{a18}) is
difficult to solve
in the general causal case $\beta>0$.
However, we can find a perturbative solution in the case of small
relaxational effects, and we can also find an exact solution
in the case
$\sigma = 0$, i.e. when the mean collision time is constant. This
assumption may be reasonable over short periods of time.

\subsection*{3.1 Small relaxational effects}

For a qualitative estimate of how relaxation effects change the
predictions of the quasi-stationary theory, we consider the case of
small relaxation parameter $\beta\gamma/\alpha$, and
solve equation (\ref{a18}) perturbatively. We write
\[
T = \tilde{T} + \left({\beta\gamma\over\alpha}\right)S + 
O\left[\left({\beta\gamma\over\alpha}\right)^2\right]  \,,
\]
where the quasi-stationary temperature $\tilde{T}$
is given by equations (\ref{roy1})--(\ref{roy3}).
In order to make sensible comparisons, we require the
luminosities, and thus the surface temperatures, to coincide,
so that
\[
S_{_\Sigma}=0 \,.
\]
To first order, the temperature equation (\ref{a18}) gives
\[
\left(A_0S\right)^\prime+(3-\sigma){(A_0\tilde{T})^\prime\over
(A_0\tilde{T})}\,\left(A_0S\right)={2\over\gamma}
\left({\ddot{f}\over f}-2
{\dot{f}^2\over f^2}\right)\left(A_0\tilde{T}\right)^{-3}
A_0{A_0}^\prime\,.
\]
This linear equation in $A_0S$ has solution
\be
\left(A_0S\right) = -{2\over\gamma}\left({\ddot{f}\over f}-2{\dot{f}^2
\over f^2}\right)\left(A_0\tilde{T}\right)^{\sigma-3}
\int_r^{r_{_\Sigma}}
\left(A_0\tilde{T}\right)^{-\sigma}A_0{A_0}^\prime dr \,,
\label{roy5}
\ee
where $\tilde{T}$ is given in equations (\ref{roy1})--(\ref{roy3}).
Now it follows from equations (\ref{a6})--(\ref{a8}) that
\[
{\ddot f\over f}-2{\dot{f}^2\over f^2}=a{\dot f\over f^2}
\left({5\over\sqrt f}- 4\right) \,,
\]
which is always negative, since $0\leq f\leq 1$ and
$\dot f<0$. Thus we see that
\[
S>0 \,,
\]
so that small relaxational effects lead to an increase in the
temperature into the interior, consistent with the
results of Herrera and Santos (1997).

For the case of thermal neutrino transport, where $\sigma={3\over2}$,
we can give explicit analytic forms for the relaxational
temperature correction. By equation (\ref{roy1})
\[
\left(A_0\tilde T\right)^{5/2}={10\over3\alpha}{\dot f\over f}
{A_0}^{3/2}+{\cal F}(t)\,.
\]
We find ${\cal F}$ from the surface conditions via equations 
(\ref{a21}),
(\ref{a19}) and (\ref{a20})
\be
{\cal F}(t) = \left\{\left[\frac{2a^2 {A_0}^4}{\xi r^2}\frac{(1 -
\sqrt{f})}{{f}^{3}\sqrt{f}}\left(\frac{rB_0f -
2m}{A_0(B_0r)^{\prime}\sqrt{f} + 2ar{B_0}^2(1 -
\sqrt{f})}\right)^2\right]^{5/8}
- \frac{10}{3\alpha}{A_0}^{3/2} \frac{\dot
f}{f}\right\}_{_\Sigma}\,.
\label{F} \ee
Now the integral in equation (\ref{roy5}) becomes
\begin{eqnarray*}
&&\int_0^{r_{_\Sigma}}\left[{10\over3\alpha}{\dot f\over f}
{A_0}^{3/2}+F(t)\right]^{-3/5} A_0dA_0 = \\
&&\left\{{3\alpha\over 11}{f\over{\dot
f}}{A_0}^{1/2}
\left[{10\over 3\alpha}{{\dot f}\over f}{A_0}^{3/2} + {\cal F} - {\cal
F}^{2/5}
{\bf F}\left({1\over 3},{3\over 5}, {4\over 3}, -{10\over 3\alpha{\cal
F}}{{\dot f}\over f}{A_0}^{3/2}\right)\right]
\right\}_0^{r_{_\Sigma}}\,,
\end{eqnarray*}
where ${\cal F}$ is given by equation
(\ref{F}) and ${\bf F}(a, b, c, z)$ is
the hypergeometric function.
Thus, collecting the above results, we have the explicit analytic
form for the causal temperature correction in the case of thermal
neutrino transport with small relaxation timescale:
\[
S= {2\over\gamma}\left({\ddot{f}\over f}-2{\dot{f}^2
\over f^2}\right)\left(A_0\tilde{T}\right)^{\sigma-3} 
\left\{{3\alpha\over 11}{f\over{\dot
f}}{A_0}^{1/2}
\left[{10\over 3\alpha}{{\dot f}\over f}{A_0}^{3/2} + {\cal F} - {\cal
F}^{2/5}{\bf F}\left({1\over 3},{3\over 5}, {4\over 3}, 
-{10\over 3\alpha{\cal F}}
{{\dot f}\over f}{A_0}^{3/2}\right)\right]
\right\}_0^{r_{_\Sigma}}\,.
\]

\subsection*{3.2 Constant collision time approximation}

For $\sigma = 0$, equation (\ref{a18}) integrates to give
\be  \label{aa18}
T^4 = \frac{{\cal F}(t)}{{A_0}^4} +  4\frac{\beta}{\alpha
{A_0}^2}\left[\frac{\ddot f}{f} - 2\frac{{\dot f}^2}{f^2}\right] +
\frac{8}{3\alpha {A_0}}\frac{\dot f}{f}\,,
\ee
where ${\cal F}(t)$ is a function of integration.
Using equations
(\ref{aa18}), (\ref{a19}), (\ref{a20}) and (\ref{a21}) we
obtain
\begin{eqnarray*}
{\cal F}(t) &=& \left\{\frac{2a^2 {A_0}^4}{\xi r^2}\frac{(1 -
\sqrt{f})}{{f}^{3}\sqrt{f}}\left[\frac{rB_0f -
2m}{A_0(B_0r)^{\prime}\sqrt{f} + 2ar{B_0}^2(1 - \sqrt{f})}\right]^2
\right. \\
&&{}\left. -
4\frac{\beta {A_0}^2}{\alpha} \left(\frac{{\ddot f}}{f} - 
2\frac{{\dot
f}^2}{f^2}\right) - \frac{8}{3\alpha}{A_0}^3 \frac{\dot 
f}{f}\right\}_{_\Sigma}\,.
\end{eqnarray*}
Hence we can finally write the temperature as
\beq \label{fff}
T^4 &=& -\frac{8a^2 \beta}{\alpha f^3}\left[\frac{1}{{A_0}^2} -
\frac{{({A_0}^2)_{_\Sigma}}}{{A_0}^4}\right](5 + 4f - 9\sqrt{f}) -
\frac{16a}{f\sqrt{f}}(1 - \sqrt{f})\left[\frac{1}{{A_0}} -
\frac{{({A_0}^3)_{_\Sigma}}}{{A_0}^4}\right] \n \\ 
&&{}+ \left\{\frac{2a^2}{\xi}{A_0}^4\frac{(1 -
\sqrt{f})}{{f}^{3}\sqrt{f}r^2}\left[\frac{rB_0f -
2m}{A_0(B_0r)^{\prime}\sqrt{f} + 2ar{B_0}^2(1 -
\sqrt{f})}\right]^2\right\}_{_\Sigma} \,.
\eeq

If we set $\beta = 0$ then (\ref{fff}) becomes
\beq \label{fff1}
{\tilde T}^4 &=& - \frac{16a}{f\sqrt{f}}(1 - 
\sqrt{f})\left[\frac{1}{{A_0}} -
\frac{{({A_0}^3)_{_\Sigma}}}{{A_0}^4}\right] \n \\ 
&&{}+ \left\{\frac{2a^2}{\xi}{A_0}^4\frac{(1 -
\sqrt{f})}{{f}^{3}\sqrt{f}r^2}\left[\frac{rB_0f -
2m}{A_0(B_0r)^{\prime}\sqrt{f} + 2ar{B_0}^2(1 -
\sqrt{f})}\right]^2\right\}_{_\Sigma}    \,,
\eeq
where ${\tilde T}$ is the noncausal temperature. This is exactly the
temperature obtained by de Oliveira and Santos (1985) when
considering shear-free stellar models in the noncausal theory. Our
expression for the temperature in (\ref{fff}) differs by the 
additional
term involving $\beta$.

\section*{4 Conclusions}

Using a very simple shearfree stellar metric, we have investigated
relaxational effects on the temperature in a radiating collapsing star
in the non-perturbative regime relative to the static limit.
Our results are in this sense a non-perturbative (though restrictive,
due to the simplicity of the metric) complement to the perturbative
results of Herrera and Santos (1997), whose model necessarily remains
close to the static limit.

In the case of small relaxation time-scale 
but possibly large
deviations from staticity, we were able to confirm in section 3.1
the increase in interior temperature that results from
causal heat transport. In other words, heat transport that respects 
relativistic causality requires that the interior of a star must be
hotter than predicted by quasi-stationary, non-causal transport, for
a given luminosity. We also found the explicit analytic form of
the temperature correction in terms of hypergeometric functions.

In the case of unrestricted relaxation timescale (section 3.2), 
we were able
to find analytic forms only for the constant collision-time
regime.
A detailed analysis of the causal temperature $T$ in (\ref{fff}) is 
not
easy because of the presence of the arbitrary functions $A_0$ and 
$B_0$.
However it is possible to make a few general qualitative statements
about the behaviour of $T$. It is easy to see that the causal 
and noncausal temperatures are different in the interior of the star. 

In particular we note that the causal temperature is higher
for a given luminosity than the
noncausal temperature for each interior point of the star since 
\[ 
4f - 9\sqrt{f} + 5 > 0 \,. 
\]
Here we have used the fact that $0 < f < 1$ and 
the result 
\[
1 - \frac{({A_0}^2)_{_\Sigma}}{{A_0}^2} < 0 \,,
\]
which follows since
${A_0}^{\prime} > 0$ in the interior of the star. The parameter
$\beta$ determines the deviation from the noncausal temperature 
${\tilde T}$. For small values of $\beta\gamma/\alpha$, 
the causal temperature is similar to
the noncausal temperature ${\tilde T}$, but as relaxational effects 
grow
for larger values of $\beta\gamma/\alpha$,
$T$ and ${\tilde T}$ deviate
substantially.

The causal and noncausal temperatures
coincide on the surface of the radiating star:
\[
(T)_{_\Sigma} = ({\tilde T})_{_\Sigma} \,.
\]
By equation 
(\ref{fff}) we note that the surface temperature $T_{_\Sigma}$
becomes zero at $t = -\infty$ and at the time of the
formation of the horizon given by 
\[
f = \frac{2m}{(rB_0)_{_\Sigma}}\,.
\] 
This is also the case for the noncausal theory, from
equation (\ref{fff}).
Note that
\be \label{ccc}
\kappa(T)({A_0}T)^{\prime} - \kappa({\tilde T})({A_0}{\tilde 
T})^{\prime} =
-(B_0f\tau){\dot q} \,.
\ee
Using equations
(\ref{a15}) and (\ref{a16}) we can rewrite equation (\ref{ccc}) as
\be \label{ccc1}
\left({\cal T}^{4 - \sigma}\right)^{\prime} - \left({\tilde{\cal 
T}}^{4 -
\sigma}\right)^{\prime} = - \frac{4 - \sigma}{\alpha}\left({A_0}^{3 -
\sigma} B_0 f
\tau\right){\dot q} \,,
\ee
where we have defined ${\cal T} = {A_0}T$ and ${\tilde{\cal T}} =
{A_0}{\tilde T}$. We can immediately see that the relative spatial 
gradient of
${\cal T}$ is greater than that of ${\tilde {\cal T}}$, since
$f$ and $\dot q$ are positive.
Equation (\ref{ccc1}) is similar to the result of 
Govender et al. (1998).
However in that case the acceleration was vanishing and 
the
model had a Friedmann limit. In the present model there is nonzero
acceleration and the model has a static limit at $t = -\infty$.

To conclude we observe that the line element (\ref{x1}) used
by Herrera and Santos is more general than (\ref{a1a}). However,
we were
in a position to investigate the evolution of the temperature profile
analytically. Using the causal
transport equation for heat flow we were able to demonstrate the
role played by the relaxation time in radiative gravitational
collapse, providing a non-perturbative support for the 
results of Herrera and Santos. Looking
to the future, a more realistic analytic model of dissipative
gravitational collapse would need to incorporate the effects of
viscosity and shear. It would be interesting to investigate the
relaxational effects of such a model especially during the late
stages of collapse when the temperatures can be of the order of
$10^9$K
(Shapiro and Teukolsky, 1983).


\end{document}